\documentclass[twocolumn,prb,superscriptaddress,showpacs]{revtex4}

\usepackage{graphicx}
\usepackage{amsmath}
\usepackage{amstext}
\usepackage{amssymb}
\usepackage[colorlinks,citecolor=blue]{hyperref}
\usepackage{graphicx}
\usepackage{amsmath}
\usepackage{amstext}
\usepackage{longtable}
\usepackage{amssymb}
\usepackage{amsfonts}
\usepackage{longtable,booktabs}
\usepackage{hyperref}\usepackage{url}
\usepackage{ulem}

\usepackage{amsbsy}
\usepackage{dcolumn}
\usepackage{amsthm}
\usepackage{bm}
\usepackage{multirow}
\usepackage{hyperref}
\hypersetup{
    colorlinks=true,
    linkcolor=blue,
    filecolor=magenta,
    urlcolor=cyan,
}

\usepackage{mathrsfs}
\usepackage{amsfonts}
\usepackage{amsbsy}
\usepackage{dcolumn}
\usepackage{bm}
\usepackage{multirow}
\usepackage{color}
\def\Z{\mathbb{Z}}

 \usepackage{graphicx}
    \usepackage{amssymb}
    \usepackage{amsthm}
    \usepackage{mathtools}

\usepackage{float}

      \theoremstyle{remark}

\begin{document}

\title{{\fontfamily{ptm}\fontseries{b}\selectfont Topological Superconductivity by Doping Symmetry-Protected Topological States}} 

\author{Shang-Qiang Ning}
\affiliation{Institute for Advanced Study, Tsinghua University, Beijing, China, 100084}
\author{Zheng-Xin Liu}
\email{liuzxphys@ruc.edu.cn}
\affiliation{Department of Physics, Renmin University of China, Beijing, China, 100872}
\author{Hong-Chen Jiang} 
\email{hcjiang@stanford.edu}
\affiliation{Stanford Institute for Materials and Energy Sciences, SLAC and Stanford University, 2575 Sand Hill Road, Menlo Park, CA 94025, USA}

\begin{abstract}  
We propose an exotic scenario that topological superconductivity can emerge by doping strongly interacting fermionic systems whose spin degrees of freedom form bosonic symmetry protected topological (SPT) state. Specifically, we study a 1-dimensional (1D) example where the spin degrees of freedom form a spin-1 Haldane phase. {Before doping, the charge and spin degrees of freedom are both gapped.} Upon doping, {the charge channel becomes gapless and is described by a  $c=1$ compactified bosonic conformal field theory (CFT), while the spin channel remains gapped and still form a bosonic SPT state. Interestingly,} an instability toward $p$-wave topological superconductivity is induced coexisting with the symmetry protected spin edge modes that are inherited from the Haldane phase. This scenario is confirmed by density-matrix renormalization group simulation of a concrete lattice model, where we find that topological superconductivity is robust against interactions. We further show that by stacking doped Haldane phases an exotic 2D anisotropic superconductor can be realized, {where the boundaries transverse to the {chain-}direction are either gapless or spontaneously symmetry-broken due to the Lieb-Schultz-Mattis (LSM) anomaly.} 
\end{abstract}
 \maketitle

 The interplay between symmetry and topology plays an important role in modern condensed matter physics, and leads to the discovery and classification of various types of exotic (gapped) phases of matter, one of which is the symmetry-protected topological (SPT) phases\cite{wengu09}. Characteristic SPT phases include the topological insulators \cite{kanemele2005, TI4, TI6}, Haldane phase\cite{haldane83_0, haldane83} and bosonic integer quantum Hall liquids\cite{biqh1}. In SPT phases, the bulk excitations are gapped and unfractionalized while the edge states can be either gapless or topologically ordered if the symmetry is not broken\cite{CZX_model}.

Bosonic SPT (bSPT) phases are (partially) classified by group cohomology theory\cite{chenguliuwenscience,chenguliuwenprb}. However, the classification of fermionic SPT (fSPT) phases is rather involved. Non-interacting fermionic topological phases are classified by K-theory\cite{tenfoldways,kitaev09ktheory}. When interactions are turned on, the classes of fSPT phases will generally be changed. A typical example is 1D time-reversal invariant topological superconductors, where interactions reduce the non-interacting $Z$ classes into $Z_8$\cite{Fidkowski10, fidkowski11}. A systematic classification of interacting fSPT phases is relatively difficult while remarkable progresses have been made recently\cite{wengu14, mengchenjie18, wanggu17, wanggu18}. Roughly speaking, interacting fSPTs can be divided into three types\cite{chenjie16}: (I) those are adiabatically connected to non-interacting fermion SPT phases; (II) those are adiabatically connected to bosonic SPT phases; (III) those can only be realized in interacting fermionic systems. While type-I and type-III fSPT phases only exist in pure fermionic systems, the Type-II phases can be obtained by freezing the fermionic charge degrees of freedom to form Mott insulators (or by enforcing the fermions to tightly bind into bosonic molecules) and then letting the bosonic spin (or molecule) degrees of freedom form a bSPT\cite{sq15}. Some of the type-II fSPT phases can also be obtained by stacking a band insulator with a bSPT.
In other words, such type of phases can be realized by embedding bSPT phase in fermionic system and can be called bSPT-embedded phases\cite{chenjie16}.

In this paper, we investigate the physical consequence by doping holes into the Type-II fSPT insulators. We expect that the holes in such system can exhibit nontrivial charge behavior, which drives the system into a superconductor. 
This is illustrated by a concrete example, where the embedded bSPT is the spin-1 Haldane-phase whose dangling spin-1/2 edge states are protected by time reversal, or $SO(3)$ (or its subgroup $\Z_2\times \Z_2$) symmetry \cite{wengu09, ES_of_1DSPT_pollmann_prb}.  The way we embed the spin-1 Haldane phase into nontrivial fSPT is to introduce three species of spinless fermions $\psi_x, \psi_y, \psi_z$ on each site\cite{sq15}. We show that the Haldane phase at 1/3 filling has a topological superconducting instability upon doping. This scenario is confirmed by large-scale density-matrix renormalization (DMRG) \cite{dmrg} simulations. We further conclude that the superconducting instability can be turned into a $p$-wave superconductor by coupling the 1D wires to from a 2D lattice.

{\it Continuous field theory. } Similar to \cite{sq15}, we consider a fermion ladder with three species $\psi_x, \psi_y, \psi_z$, whose Hamiltonian is $H=H_0 + H_I${.}   Here $H_0$ is the non-interacting part $H_0=-t\sum_{\alpha,i} \psi_{\alpha, i}^\dag  \psi_{ \alpha,i+1} + h.c.$ and $H_I$ is the interacting part which will be specified later. Different $H_1$ can realize different phases. The system contains time reversal symmetry, global $U(1)$ symmetry generated by the total fermion charge $\hat n=\sum_{\alpha,i} \psi^\dagger_{\alpha,i} \psi_{\alpha,i}$ and $SO(3)$ symmetry(or one of its subgroup $\Z_2\times \Z_2$) which mixes the three species. Assuming that the interaction is not strong, we can treat $H_I$ as perturbations and use bosonization technique to analyze the physical consequences. By adjusting the interactions $H_I$ and the filling, several interesting phases can be obtained.

The unperturbed part $H_0$ is described by three isotopic Luttinger liquid  with action
\begin{align}\label{S0}
H_0=\frac{v_F}{2\pi}\int dx \text{{$\sum_{\alpha=x,y,z}$}} (\partial_x \phi_\alpha)^2+(\partial_x \theta_\alpha)^2
\end{align}
where $\phi_{\alpha}$ are the bosonized fields and  $\theta_{\alpha}$ are their dual fields which satisfy $[\partial_x\phi_\alpha(x),  \theta_\beta(x')]=i\pi  \delta_{\alpha, \beta}\delta(x-x')$. The spatial derivative of $-\pi\phi_{\alpha}$ (or $\pi\theta_\alpha$) is the density(or current) of the fermion $\psi_{\alpha}$.  We transform the original bases of these fields into one charge channel and two spin channels, namely, $\phi_c=\sqrt{1/3} ( \phi_x +\phi_y+\phi_z)$, $\phi_{s_1}=\sqrt{2/3} \phi_x+\sqrt{1/6} (\phi_y+\phi_z)$ and $\phi_{s_2} = \sqrt{1/2} (\phi_z-\phi_y)$. 
In this way, $\partial_x \phi_c, \partial_x \phi_{s_{1,2}}$ stand for the total charge and spin densities, respectively. For later convenience, we define the fields $\Phi_{1}=\phi_{x}-\phi_y$ and $\Phi_2=\phi_y-\phi_z$ as well, which do not involve the total charge degree of freedom. The dual fields $\theta_c,\theta_{s_{1,2}}$ and $\Theta_1,\Theta_2$ fields are defined similarly.

The Luttinger liquid in (\ref{S0}) can be gapped out by condensing some bosonic fields (with suitably chosen $H_I$). There are several ways, for instance, one can condense (1) $\phi_c$ and $\phi_{s_{1,2}}$, or (2) $\phi_c$ and $\theta_{s_{1,2}}$, or (3) $\theta_c$ and $\phi_{s_{1,2}}$, or (4) $\theta_c$ and $\theta_{s_{1,2}}$. Different condensation patterns result in different phases. To obtain a bSPT embedded phase, we require that the spin degrees of freedom are gapped in a nontrivial way. To achieve this, we condense the $2\Theta_1,2\Theta_2$ fields\footnote{The condensation of $2\Theta_{1,2}$  is obviously invariant under $\Z_2\times \Z_2$, a subgroup of $SO(3)$. Though  the SO(3) symmetry in not apparent in Abelian bosonization,  this subgroup can also protect the same Haldane phase. We further argue that this condensation can indeed be SO(3) invariant since it can come from some SO(3) invariant interactions.}.
The physical meaning of $2\Theta_1,2\Theta_2$ is given by $e^{i2\Theta_1(x)} \sim \Delta_x^\dag(x) \Delta_y(x)$ and $e^{i2\Theta_2(x)} \sim \Delta_y^\dag(x) \Delta_z(x)$, where $\Delta_ \alpha=\psi_{R\alpha(x)} \psi_{L\alpha} (x+a)$, $\alpha=x,y,z$. The charge gap can be opened by Umklapp process which scatters the three left moving fermions into the three right moving fermions and vice versa. This process keeps the $U(1)$ and $SO(3)$ symmetry unbroken, but it requires that the fraction of the particle number filling is commensurate, such as 1/3 \cite {lecheminant_SU(n)_Hubbard, SU(n)_huubard_DMRG}. As illustrated in Appendix.\ref{S1}, the condensation of $\phi_c$ and $2\Theta_1$,$2\Theta_2$ at 1/3 filling indeed results in fully gapped ground states whose degeneracy is 4 under open boundary condition, and the edge modes carry nontrivial projective representation of the  $SO(3)$. This state is adiabatically connected to the spin-1 Haldane chain.

Once the system is slightly doped away from 1/3 filling, the Umklapp process will be suppressed and hence the $\phi_c, \theta_c$ fields become gapless {and are characterized by the $c=1$ compactified bosonic CFT
\begin{align}
H_c= \frac{1}{2\pi} \int  dx [u_c K_c(\partial\theta_c)^2 + \frac{u_c}{K_c} (\partial\phi_c)^2 ], 
\label{eqn:charge_dope}
\end{align}
where $u_c$ is the renormalized fermi velocity and  $K_c$ the radius parameter. In the non-interacting limit, $u_c=v_F$ and $K_c=1$.} Meanwhile, due to spin-charge separation, the condensation of $2\Theta_{1,2}$ is unaffected and the spin channels are still gapped leaving the  spin-edge modes at open boundaries untouched. In this case, the dominant correlation function  is of the $SO(3)$-singlet pairing $\Delta_{Sc}=\sum_{\alpha=x,y,z}\Delta_ \alpha$, which has odd-parity under fermion exchanging(see Appendix.\ref{S2}). In bosonized fields, this quantity is written as 
$\Delta_{Sc} =\frac{e^{i\frac{2}{\sqrt{3}} \theta_{c}} }{2\pi a }  [e^{i\sqrt{\frac{8}{3}}\theta_{s_1}}+2e^{i\sqrt{\frac{2}{3}}\theta_{s_1} }\text{cos}( \sqrt{2} \theta_{s_2})] + ... $, where the cut-off $a$ is the lattice constant. Since $\theta_{s_1}$ and $\theta_{s_2}$ are condensed and $\theta_c$ is gapless in the doped fermionic Haldane phase, $\langle \Delta_{Sc}(r+r_0)\Delta_{Sc}(r_0) \rangle \propto {1/ r^{2\over3K_c}}$,  $K_c$ is the Luttinger parameter of the charge channel. 

On the other hand, if we condense the $2\Phi_1$ and $2\Phi_2$ fields, then the degenerate spin-edge modes go away, and a trivial gapped phase is obtained. In this case,  under light doping $\phi_c$ is gapless {and is still characterized by (\ref{eqn:charge_dope}),} where all superconducting correlation functions are exponentially decaying while the total particle density wave correlation becomes power-law decaying (see Appendix.\ref{S2}). 

Above we have shown that the condensation of $\phi_c, 2\Theta_{1}, 2\Theta_{2}$ at 1/3 filling gives rise to the Haldane phase. When doped away from the commensurate filling, the system shows an instability toward an $SO(3)$-invariant odd-parity superconductivity. Above scenario can be realized by adding several simple interactions to the system. A natural interaction is the antiferromagnetic Heisenberg exchange. To drive the system into a Mott insulator, a Hubbard interaction is also plausible. Notice that in above discussion we have assumed that the interactions are not strong, otherwise it may cause obstacle to the hopping of charge and lead to phase separation (see below for more details).

\begin{figure}
\includegraphics[width=\linewidth, height=5cm] {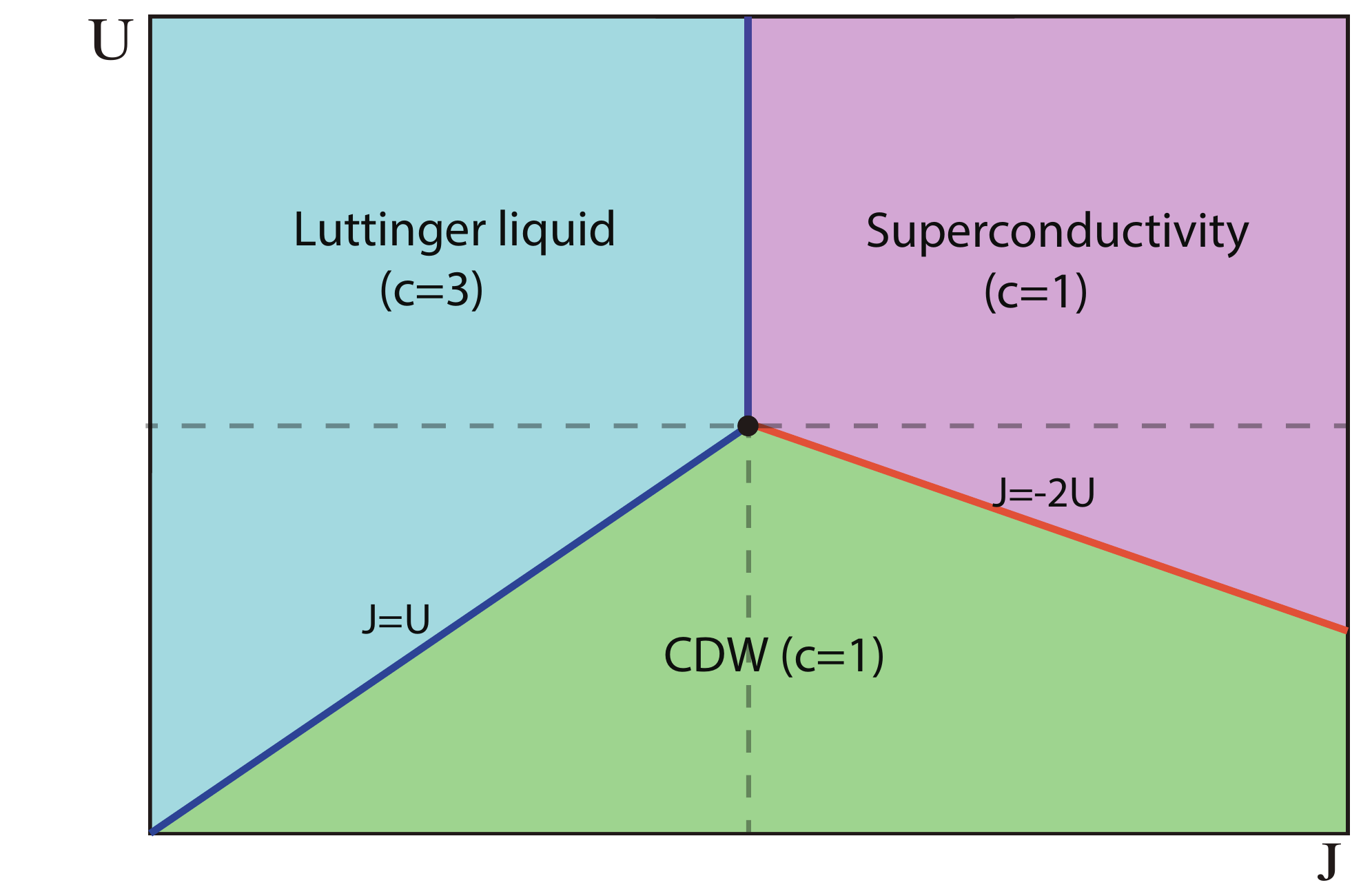}
\caption{Bosonization phase diagram with light doping for Hamiltonian with interaction (\ref{eqn:interaction}). ${c}$ represents the central charge.  { The charge channel is always gapless with ${c}=1$. }The dominant instability of the red regime is superconductivity which is $\Delta_{Sc}=\sum_\alpha \psi_\alpha(x) \psi_\alpha(x+a)$ while that of the green regime is density wave instability which is $n(x)=\sum_\alpha n_{\alpha}(x)$. The Luttinger liquid is gapless both in spin and charge channels.}\label{Fig:bosonize_phase_diagram}
\end{figure}

{\it Lattice model.} The discussion above leads us to study the following interaction(together with the aforementioned isotopic hopping terms $H_0$),
\begin{eqnarray}
H_I= J\sum_i \mathbf{S}_i\cdot \mathbf{S}_{i+1}  + U\sum_i (n_i-1)^2
\label{eqn:interaction}
\end{eqnarray}
where $n_i=\sum_\alpha \psi_{\alpha,i}^\dag\psi_{\alpha,i}$ and $S_i^\alpha=i \epsilon^{\alpha\beta \gamma} \psi_{\beta,i}^ \dagger \psi_{\gamma,i}$ (these spin operators satisfy the $su(2)$ commutation relation {if the constraint $n_i=1$ is satisfied}). When $J \neq 0$, the model has $SO(3)$ symmetry which is enlarged to $SU(3)$ in the case $J=0$ (In the following, we assume $J>0$). Now we fix the filling at $1/3-\delta$ where $\delta$ denotes the doping density. 

{\it Strong coupling limit.} When the filling is $1\over3$ and when $U\gg J>t>0$, the system tends to occupy one fermion per site. On-site charge fluctuations are strongly suppressed and the model becomes a purely spin system, i.e., the spin-1 Heisenberg model \cite{sq15} which belongs to the Haldane phase. Perturbations can be added to the system without destroying the Haldane phase as long as time reversal or $Z_2\times Z_2$ symmetry is unbroken. However, in the present work we don't discuss such perturbations in details.

\begin{figure}
  \includegraphics[width=\linewidth] {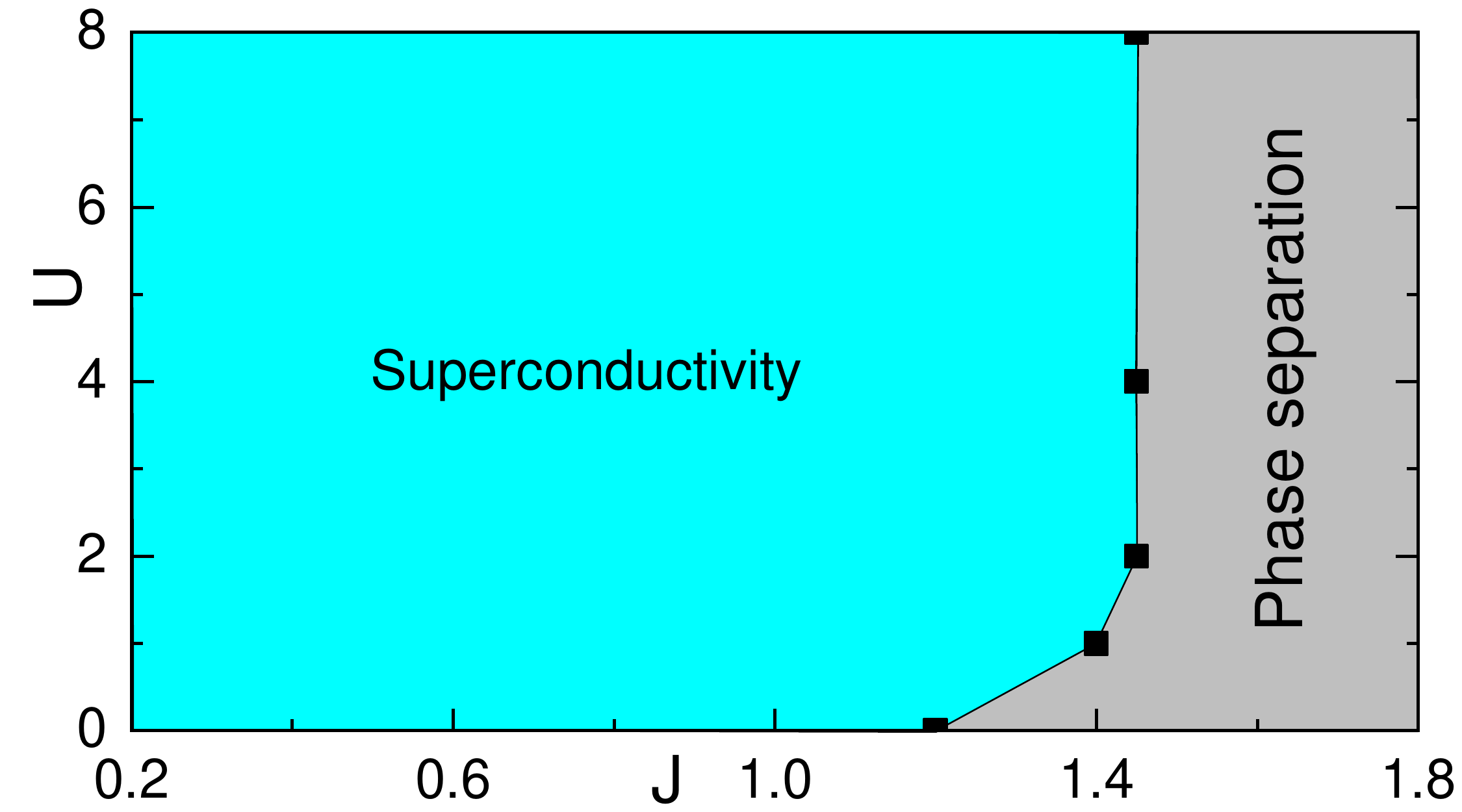}
  \caption{(Color online) Ground state phase diagram of the model in Eq.(\ref{eqn:interaction}) at doping level $\delta=5\%$ as a function of $U$ and $J$.}\label{Fig:Phasediagram}
\end{figure}

{\it Weak coupling limit.} When the interaction strength is much smaller than $t$, we can treat the interactions perturbatively. Then above Hamiltonian decouples into two channels in bosonization terminology, i.e., the charge channel and spin channel \footnote{ One bosonization study of a model similar to (\ref{eqn:interaction}) in Ref.\cite{berg_18_1DTSC} is closely related to this but also with some significant difference: filling (ref.\cite{berg_18_1DTSC} focus on $1/3$ filling and we mainly focus on doping away from $1/3$ filling), and bare interactions (the pairing interaction $V$ in Ref.\cite{berg_18_1DTSC}) are only part of the Heisenberg-like interaction $J$ in our paper.}. We first consider the charge channel,
{$
H_c'=H_c+H_u
$
where  $H_c$ is (\ref{eqn:charge_dope}) with} $K_c=1+\frac{a}{2 \pi v_F} (8\text{cos}(2k_Fa) J-8U )$ and $k_F=\pi/3-\pi \delta$ {and $H_u$ represents the Umklapp process term for $\delta=0$\cite{lecheminant_SU(n)_Hubbard, SU(n)_huubard_DMRG}}.  Without doping, the Umklapp process from higher order perturbation can cause $\phi_c$ to condense such that the charge channel opens a gap\footnote{Reference \cite{lecheminant_SU(n)_Hubbard} concluded that finite critical interaction strength is needed for the Umklapp process to open the charge gap. However, careful numerical calculation in reference \cite{ SU(n)_huubard_DMRG} suggests that the critical interaction strength is approaching to zero.}.  At not very large doping, i.e.$-1/6<\delta< 1/12$, $\text{cos}(2k_Fa)<0$ and then $K_c<1$ for positive $2U+2\text{cos}(2k_Fa)J $.  However, this process is immediately suppressed by nonzero but small $\delta$.
   
Now we consider the spin channel,
$
H_s= \frac{1}{2\pi} \int  dx \sum_{\beta=s_1,s_2} [u_\beta K_\beta (\partial\theta_\beta)^2+\frac{u_\beta}{K_\beta}(\partial\phi_\beta)^2 ]+
g[\text{cos}(\sqrt{2}\phi_{s_2}-\sqrt{6}\phi_{s_1})+\text{cos}(\sqrt{2}\phi_{s_2}+\sqrt{6}\phi_{s_1})
+\text{cos}(2\sqrt{2}\phi_{s_2})] + h[\text{cos}(\sqrt{2}\theta_{s_2}-\sqrt{6}\theta_{s_1})
+\text{cos}(\sqrt{2}\theta_{s_2}+\sqrt{6}\theta_{s_1})+\text{cos}(2\sqrt{2}\theta_{s_2})], 
$
where $K_{s_1}=K_{s_2}=1+\frac{a}{2 \pi v_F} (4U-4\text{cos}(2k_Fa) J ) =K_s$. The one-loop RG equations are
\begin{align}
\frac{dy_g}{dl}=(2-\Delta_g)y_g-y_g^2, \ \ \ 
\frac{dy_h}{dl}=(2-\Delta_h)y_h-y_h^2
\end{align}
where we have set $y_{g,h}=g(h)/2\pi v_F$ and $\Delta_g=2K_s$ and $\Delta_h=2/K_s$ are the scaling dimention of $g$ and $h$ interaction terms. 

For simplicity we assume the doping is small such that $k_F\simeq \pi/3$, $\text{cos}(2k_F a)\simeq -1/2$. (For general doping $\delta$ satisfying $-1/6<\delta<1/12$, the following discussion is still valid). From the perturbative RG flow of above equations we obtain the whole phase diagram shown in Fig.\ref {Fig:bosonize_phase_diagram}. We are specially interested in the superconductivity regime with {$J>0$ and} $2U+J>0$ for light doping (the red area in Fig.\ref {Fig:bosonize_phase_diagram}). In this case $\Delta_g>2, \Delta_h<2$, thus at the tree level, $\text{cos}(2\phi_\alpha-2\phi_\beta)$ are irrelevant while  $\text{cos}(2\theta_\alpha-2\theta_\beta)$ are relevant. Under one-loop approximation, the above RG-equations indicates that $y_g$ flows to zero while $y_h$ flows to strong coupling limit. Here the initial values of $g$ and $h$ are $4U-4J$ and $-6J$, respectively.  On the other hand, in the regime $U>J$,$2U+J<0$ and $U<0$ (i.e. the green area in Fig.\ref{Fig:bosonize_phase_diagram}), then $y_g$ flows to the strong coupling limit. Notice that $y_h$ ($y_g$) going to strong coupling limit indicates the condensation of $2\Theta_{1,2}$ ($2\Phi_{1,2}$). In the rest regime, nothing is condensed and the system is still a Luttinger liquid.

\begin{figure}
  \includegraphics[width=\linewidth]{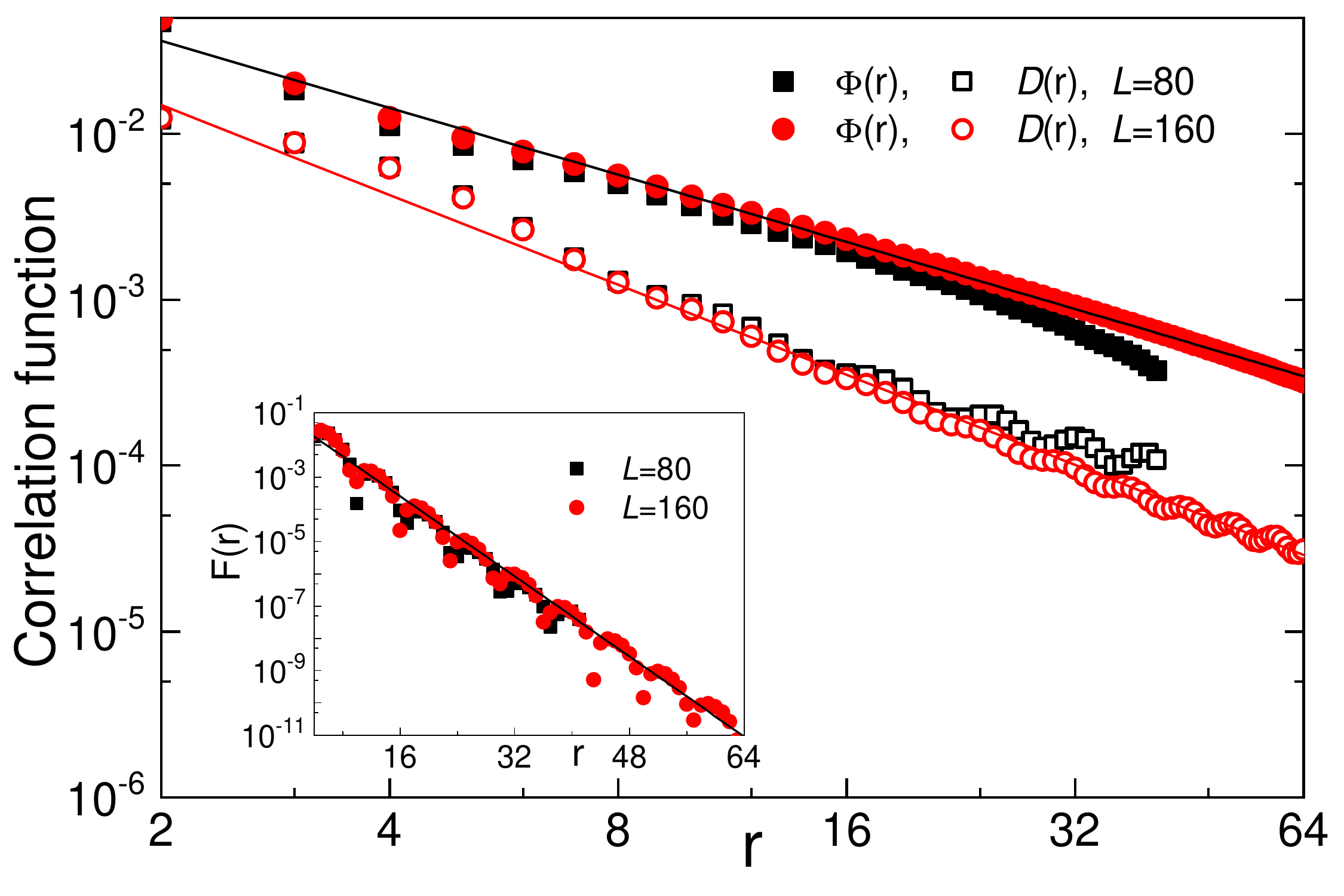}
  \caption{(Color online) Superconducting $\Phi(r)$ and density-density $D(r)$ correlation functions for the model in Eq.(\ref{eqn:interaction}) at doping level $\delta=5\%$ with $U=4$ and $J=1.0$ in a double-logarithmic plot. The solid lines are fits to $\Phi(r)\sim r^{-K_{sc}}$ (black) and $D(r)\sim r^{-K_c}$ (red), respectively. Inset is the spin-spin correlation function $F(r)$ in semi-logarithmic scales for the system, and the solid line is the fit to $F(r)\sim e^{-r/\xi_s}$.}\label{Fig:CorFunc}
\end{figure}

{\it Numerical results.} To verify above field theory analysis, now we determine the ground state phase diagram of the lattice model (\ref{eqn:interaction}) by extensive DMRG\cite{dmrg} simulations. We focus on system with open boundary condition. The total number of sites is $N=3L$, where $L$ is the length and 3 corresponds to the three-leg ladder. The number of fermions would be $N_e=L$ at $1/3$ filling and the doped hole concentration is defined as $\delta=N_h/N$, where $N_h=L-N_e$ is the number of doped holes which equals $0$ at $1/3$ filling. In our calculation, we focus on the hole concentration $\delta=5\%$. We set $t=1$ as an energy unit and keep up to $m=4000$ number of states in each DMRG block with truncation error $\epsilon < 10^{-8}$ for system length up to $L=240$. This leads to excellent convergence for our results when extrapolated to $m=\infty$ limit.

Our primary results are summarized in the phase diagram in Fig.\ref{Fig:Phasediagram} at doping level $\delta=5\%$.\footnote{We have also studied the $J<0.2$ region using DMRG. While our results are consistent with those of $J\geq 0.2$ region with superconductivity, it has visually bigger finite size effect. Hence, we will only show our results in the $J\geq 0.2$ region in this work.} Two distinct phases, including the superconducting (SC) phase and phase separation (PS) with central charge $c=1$ and $0$ respectively (see Appendix.\ref{S3}), are found with the increase of $J/t$.   The bulk of the phase diagram is occupied by the former one which is dominated by the $SO(3)$-singlet odd-parity superconducting correlation. While in the region of phase separation, the doped holes are accumulated in a small region close to the open ends while the bulk of the system is still in the fully gapped SPT phase.

Here we are mainly interested in the superconducting region, where an example for a characteristic set of parameters $J=1.0$ and $U=4$ in Fig.\ref{Fig:CorFunc}. The spin-spin correlation function $F(r)=\langle \mathbf{S}_i\cdot \mathbf{S}_{x+r}\rangle \sim e^{-r/\xi_s}$ decays exponentially, where $r$ is the distance between two sites and $\xi_s$ is the correlation length, indicating that the spin is gapped and disordered. On the contrary, the density-density correlation, defined as $D(r)=\langle \tilde{n}(x)\tilde{n}(x+r)\rangle$ where $\tilde{n}=n_i-1$, is power-law decaying $D(r)\sim r^{-K_c}$ (see Fig.\ref{Fig:CorFunc}), indicating that the charge mode is gapless which is in the $c=1$ Luttinger liquid (see Appendix.\ref{S3}). Furthermore, we find that the correlation of the previously mentioned $SO(3)$-singlet pairing operator $\Delta_{\rm Sc}$, namely $\Phi(r)=\langle \Delta_{\rm SC}^+(x) \Delta_{\rm SC}(x+r)\rangle$, also decays as power-law $\Phi(r)\sim r^{-K_{sc}}$. Interestingly, our results suggest that $\Phi(r)$ is the dominant correlation as $K_{sc}<K_c$. For instance, $K_{sc}\sim 1.34$ and $K_c\sim 1.80$ for the results in Fig.\ref{Fig:CorFunc}. We have also calculated other cases with $J=0.3$, $U= 4$ and  $J=1.0 $, $U=0.3$, and found that the exponents $K_{sc}=0.82, 1.73$ are  smaller than those of $K_{c}=1.79,1.77$, respectively. These results verified our previous prediction that the system has a instability towards $p$-wave superconductivity\footnote{Strictly speaking, the doped state with pairing correlation has a instability toward superfluidity. We consider this phase as superconducting in the sense that it can be realized in an electron system. The spinless fermion can be considered as spin-polarized electron, and hence the Haldane chain is realized by three-leg electron ladder.}.

\begin{figure}[t]
\includegraphics[width=\linewidth]{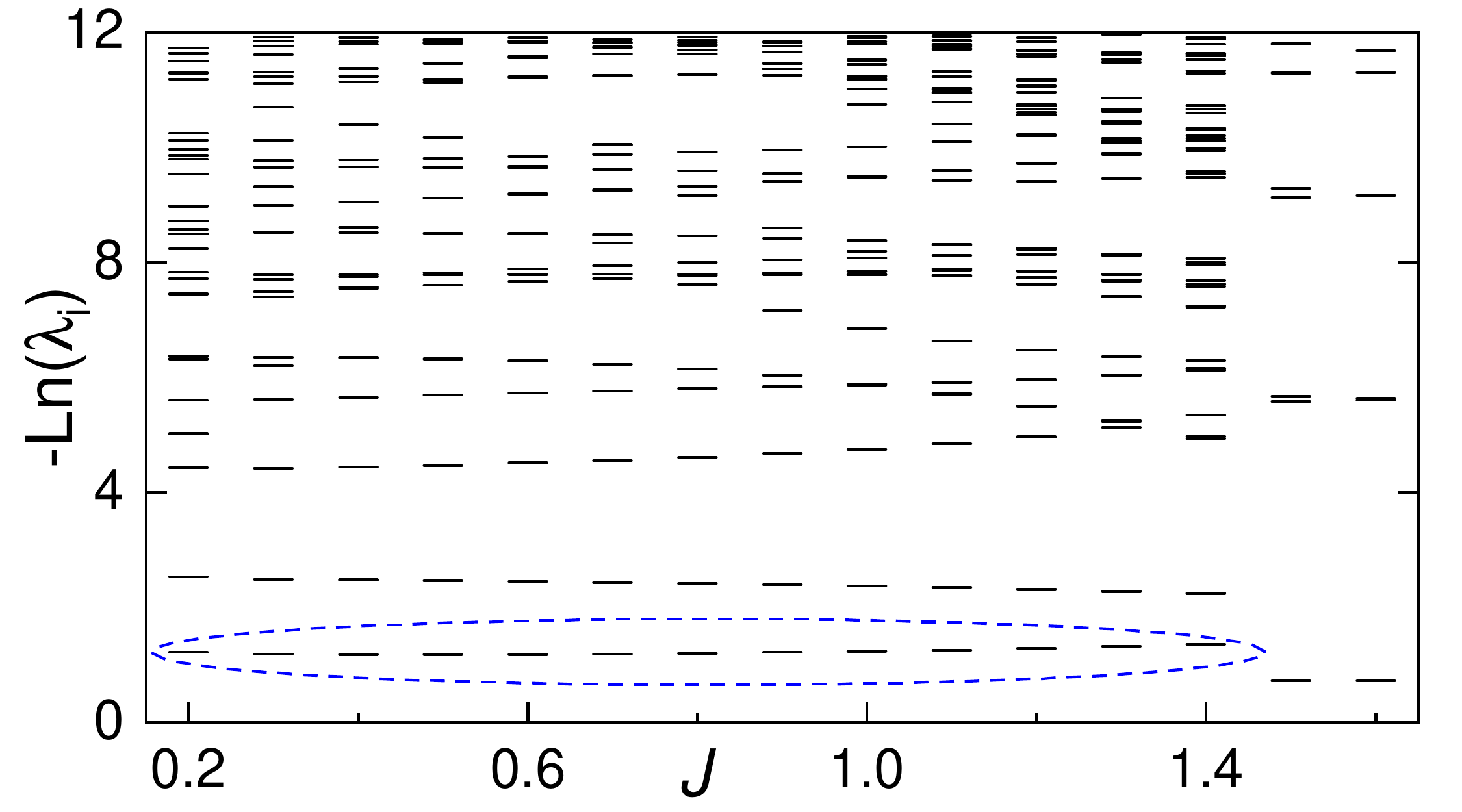}
\caption{(Color online) The 100 lowest levels of the entanglement spectrum -${\rm Ln}(\lambda_i)$ for system of length $L=160$ at $\delta=5\%$ and $U=4$ as a function of $J$, where $\lambda_i$ are the eigenvalues of the reduced density matrix. The two largest eigenvalues surrounded by the dashed blue oval are degenerate.}\label{Fig:ESpectrum}
\end{figure}

To verify the coexistence of the superconductivity and the spin-1/2 edge modes, we calculate the entanglement spectrum (ES), which is defined as the set of eigenvalues of the reduced density matrix in the ground state $\rho_A\equiv \rm{Tr}_B |\Psi\rangle \langle\Psi|$, with $A$ being a subsystem and $B$ the rest part. 
  Here we use the real space ES with OBC, where the subsystem A is chosen to be half of the chain with $N/2$ sites. The lowest level of ES in the superconducting phase is always 2-fold degenerate as shown in Fig. \ref{Fig:ESpectrum}, 
which  is consistent with the fact that there are two spin-1/2 edge modes at the open ends. { Although the charge channel is gapless whose ES whould be continuous in the thermodynamic limit, for a finite-sized system the ES contributed from the charge channel {({\it i.e.} a $c=1$ compactified bosonic CFT)} is discrete and non-degenerate. Therefore, the degeneracy in ES comes from the SPT state in the spin channel.}

When the exchange interaction is strong, the system undergoes phase separation whose central charge becomes zero. This can be understood as the competition between the kinetic energy of the holes and the exchange energy of the spins. When $J$ is large enough, the exchange wins. As a consequence, the holes accumulate at the boundary and the system forms a phase separation (see Appendix.\ref{S3}). 

{\it 2D $p$-wave superconductor by coupled wires.} Here we consider the stacking of the doped chains which have superconducting instability. We label the arrays of the chains by $j$ with $j\in Z$, and use $\{ \theta_{s_{1},j},  \theta_{s_{2},j}, \theta_{c,j} \}$ (similarly for the $\phi$ fields) to label the bosonized fields of the $j$-$th$ chain. Without inter-chain coupling, the doped Haldane-embedded chains are independent thus $\theta_{s_{1},j}, \theta_{{s_2},j}$ are condensed and $\theta_{c,j}$ are gapless for all $j\in Z$. We assume that the condensate of $\{ \theta_{s_{1},j}, \theta_{s_{2},j}\}$ is unaffected when weak inter-chain couplings are turned on. The charge channel can be gapped out in several ways, one interesting way is adding mass terms  $\text{cos}[2\sqrt{3}(\theta_{c,j} -\theta_ {c,j+1})]$ \cite{couple_wire_kane_FQHE} and  tuning them to be relevant \footnote{The scaling dimension of these mass terms is $3/K_{c,j}+3/K_{c,j+1}$ where $K_{c,j}$ is the Luttinger parameter of the charge channel in the $j$-$th$ chain.  To make these mass terms relevant, $K_{c,j}$ must be larger than 3. To this end, we have to add more interactions in each chain which only contains the forward scattering process to renormalize $K_{c,j}$. }. Resultantly, the inter-chain charge fields $\theta_{c,j}-\theta_{c,j+1}$ are locked. In the thermodynamic  limit, the total charge field $\theta_{c}=\sum_j \theta_{c,j}$ may  condense at a specific value, which leads to the spontaneous breaking of the total $U(1)$ charge symmetry and the emergence of long-range $p$-wave superconductivity . In such a state, the correlation of any inter-chain pairing order parameters e.g. $\psi_{x,j}\psi_{x,j+1}$ must be exponentially decaying since they contains the $\phi_{s_{1,2}}$ fields.  Since the $\theta_{s_{1,2},i}$ fields in each chain are still condensed, then the spin 1/2 edge modes are expected to be robust at the chain-ends. If there is no interchain interaction/hopping, these spin-1/2 edge modes would result in huge degeneracy. If the intensity of interchain interactions is much smaller than the spin gap, then these edge states will either form gapless dispersion or get the degeneracy lifted by spontaneous breaking of the translation symmetry from the {LSM theorem\cite{LSM_LSM, LSM_Meng}.} We believe that there exists a parameter window for the gapless edge, which may be described by the $SU(2)_1$ WZW model up to marginal deformations. As a consequence, we find that  such array of doping Haldane-embedded fSPT phase could lead to the coexistence of superconductivity and gapless edge modes. We can also see that the $p+ip$ superconductivity cannot be realized without breaking the condensation of $\theta_{s_{1},j}, \theta_{s_{2},j}$. We leave the realization of this 2D chiral topological superconductivity for future study.

{\it Summary and Discussion.} We study the interplay between the charge fluctuations and nontrivial topology in the spin degrees of freedom in interacting fermionic SPT phase.  As is proposed that doped spin liquids can give rise to high temperature superconductivity\cite{ANDERSON1196}, here we {demonstrate} another exotic scenario that doping the interacting fermionic SPT insulators can also lead to nontrivial superconductivity. We illustrate this possibility by a 1D example, i.e, doping the Haldane-phase-embedded fSPT away from 1/3 filling can result in nontrivial superconducting instability in certain parameter regime. We further show that stacking such superconducting wires can lead to anisotropic 2D $p$-wave superconductivity coexisting with symmetry-protected gapless edge modes in the transverse direction. 

The underlying picture is that the $U(1)$ charge conservation symmetry in the Type-II fSPT insulators, which plays no role in protecting its topological features in the spin channel, can be spontaneously broken by tuning the interactions and the filling fraction. Furthermore, the emergence of superconductivity does not affect the topological nature of the spin degrees of freedom. Therefore, an exotic superconductor is obtained. We believe that such scenario of obtaining unconventional superconductivity can be generalized in more general settings, especially in two  and higher dimensional Type-II fSPT insulators.  

An essential ingredient for the emergent superconductivity in our examples is that the charge fluctuation should not be too weak comparing to the gap in the SPT state formed by spin degrees of freedom. Although the topological properties of the SPT phases can be clearly seen in the topological limit (where the gap is set to infinity), our study suggests that away from the topological limit the competition from other degrees of freedom may also yield interesting new physics in the SPT phases.

{\it Acknowledgement.} S.Q.N. thanks Meng Cheng, Zixiang Li for insightful discussion. S.Q.N. and Z.X.L. are supported by the Ministry of Science and Technology of China (Grant No. 2016YFA0300504), the NSF of China (Grants No. 11974421 and No. 11574392), and the Fundamental Research Funds for the Central Universities and the Research Funds of Renmin University of China (No. 15XNLF19). S.Q.N. also thanks the supported from NSFC 11574172. H.C.J. is supported by the Department of Energy, Office of Science, Basic Energy Sciences, Materials Sciences and Engineering Division, under Contract DE-AC02-76SF00515. Parts of the computing for this project was performed on the Sherlock cluster.

 \bibliography{DopeSPT}


\appendix

\section{Ground state degeneracy of Haldane phase with different boundary conditions}
\label{S1}

 In this section, we will show that when $2\Theta_1$ and $2\Theta_2$ condense, the spin channel will be gapped. The ground state is four-fold degeneracy (which can be decomposed into a singlet and a triplet) under open boundary condition (OBC), while is unique under closed boundary conditions (CBC).
 
 \subsection{Open boundary condition}
Taking advantage of the spin-charge separation in the bosonized low-energy field theory, here we show that the condensation of $\theta_{s_{1,2}}$ opens a gap in the spin channel and gives rise to four-fold ground state degeneracy under OBC, which verifies that there is a spin-1/2 edge state at each boundary. 

We first make a general statement: \textit{if the fields $2\Theta_{1,2}$ condense,  the ground states of the two spin channels are four-fold degenerate under OBC which can be decomposed into a $SO(3)$ singlet and a $SO(3)$ triplet.} 
 
The condensation of $2\Theta_{1}$ and $2\Theta_2$, which is equivalent to the condensation of $\theta_{s_1}$ and $\theta_{s_2}$, suggests that the two spin channels are gapped. For simplicity, we consider the case that $2\Theta_{1,2}$ condenses at value $0$ mod $2\pi$, i.e., $\Theta_{1,2}$ take $0$ or $\pi$ mod $2\pi$, namely $|\Theta_{1,2}\rangle=|0\rangle$ or $|\pi \rangle$. With the $SO(3)$ symmetry, the fermion parity of the three species of fermions are all good quantum numbers. Since $\theta_{\alpha}$ shifts to $\theta_{\alpha}+\pi$ under the fermion parity operation $(-1)^{N_\alpha}$ of $\psi_\alpha$ fermion, only the  superposition states  of $\Theta_{1,2}$: $\Theta_{1,2}^{\pm}=|0\rangle \pm |\pi \rangle$ have well-defined quantum number of $(-1)^{N_\alpha}$, which implies that the four-fold degenerate states in the spin channels in the thermodynamic limit are the following superposition states:
\begin{align}
|\Psi_{1,2,3,4}\rangle = |\Theta_1^{\pm}\rangle \otimes |\Theta_2^{\pm}\rangle \label{ground state space}
\end{align}

Then we show that these four states can be decomposed into a $SO(3)$ singlet and a $SO(3)$ triplet. As it is not convenient to study the $SO(3)$ transformation on the bosonized fields, we consider its subgroup $\Z_2\times \Z_2$ which can protect the same nontrivial bosonic Haldane phase as $SO(3)$ symmetry.  We choose the two generators, denoted as X and Y to be the $\pi$ rotation around the $x$ and $y$ axis, i.e., under X, $\psi_x\rightarrow \psi_x$, $\psi_{y,z}\rightarrow -\psi_{y,z}$; under Y, $\psi_y\rightarrow \psi_y$, $\psi_{x,z}\rightarrow -\psi_{x,z}$. This indicates that under X, $\theta_x\rightarrow \theta_x$, $\theta_{y,z}\rightarrow \theta_{y,z}+\pi$ and under Y, $\theta_y\rightarrow \theta_y$, $\theta_{x,z}\rightarrow \theta_{x,z}+\pi$ while all $\phi$ fields remain unchanged.  Therefore, in the ground-state space expanded by $|\Psi_{1,3,4,2}\rangle$, these operators are represented by:
{\footnotesize \begin{align}
X=
\begin{pmatrix} 1 &0&0&0 \\ 0 &1&0&0 \\ 0&0&-1 &0 \\ 0&0&0&-1\end{pmatrix}, Y=
\begin{pmatrix} 1 &0&0&0 \\ 0 &-1&0&0 \\ 0&0&1 &0 \\ 0&0&0&-1\end{pmatrix},
Z=\begin{pmatrix} 1 &0&0&0 \\ 0 &-1&0&0 \\ 0&0&-1 &0 \\ 0&0&0&1\end{pmatrix}
\end{align}}where $Z=XY$.
Now we can see that  $|\Psi_1\rangle$ is invariant under $\Z_2\times \Z_2$, indicating that it is a singlet.  To show that $|\Psi_{2,3,4}\rangle$ are triplets, we make the basis transformation, $|\Psi_3\rangle =-1/\sqrt{2}(|\Psi_{1}\rangle + i |\Psi_{-1}\rangle)$, $|\Psi_2\rangle =1/\sqrt{2}(|\Psi_{1}\rangle -i|\Psi_{-1}\rangle)$, $|\Psi_4\rangle =|\Psi_{0}\rangle $. Under the new bases, the operators become
{\footnotesize \begin{align}
X'=
\begin{pmatrix} 0&0&-1 \\ 0&-1 &0 \\ -1&0&0\end{pmatrix}, 
Y'=\begin{pmatrix} 0&0&1 \\ 0&-1 &0 \\ 1&0&0\end{pmatrix},
Z'=\begin{pmatrix} -1&0&0 \\ 0&1 &0 \\ 0&0&-1\end{pmatrix}
\end{align}}
This is exactly the usual matrices of rotation operators  $X,Y,Z$ in the triplet space in the $S_z$ representation. Therefore, we conclude that these three states $|\Psi_{2,3,4}\rangle$ form a $SO(3)$ triplet. As the open chain has two ends, we can say that the four-fold degeneracy comes from the fact that at every end point, there is a dangling spin-1/2 which carries the projective representation of $SO(3)$ symmetry group and is thus protected. This statement can be further justified if we consider an additional reflection $Z_2$ symmetry seen in the density profile in Fig.\ref{density_profile}.

{In a one-dimensional open chain of length $L$,  there may be nonvanishing tunneling between these four different degenerate states which leads to the energy splitting. Since the precise expression depends on the form of Hamiltonian, we may  take the bosonized form of the lattice model to evaluate the energy splitting. For instance, we can use the same method as in \cite{Fidkowski_1106} to show that the energy splitting is proportional to $e^{-\alpha L}$ which will vanish in the thermodynamic limit.}

 \subsection{Closed boundary condition}

Contrary to the four-fold degeneracy under OBC, we now show that \textit{the ground state is non-degenerate under CBC}.

Since  $\theta_{\alpha}$($\alpha=x,y,z$) shifts by $\pi$ under the fermion parity operation $(-)^{N_\alpha}$, the two possible pinned values ($0, \pi$) of $\Theta_1$ are exchanged by either $(-)^{N_x}$ or $(-)^{N_y}$. Similarly, $\Theta_2$ would be exchanged under $(-)^{N_y}$ or $(-)^{N_z}$.    Therefore, only the following four possible combinations of states  are allowed:
\begin{equation}
|\Psi_{1,2,3,4}\rangle = ( |0\rangle \pm |\pi \rangle) \otimes (|0\rangle \pm |\pi\rangle)
\label{allowed GS}
\end{equation}
As a result, in the  ground-state space of the spin channels, we have the quantum number of parity of  $(-)^{N_x+N_y}$, $(-)^{N_x+N_z}$ and $(-)^{N_y+N_z}$ for the above four states as shown in Table.\ref{tab:parity}.

\begin{widetext}
 \begin{center}
\begin{table}[h]
    \begin{tabular}{c|c|c|c|c}
    \hline
    \hline
      & $\Psi_1=|+\rangle \otimes |+\rangle$ & $\Psi_2=|+\rangle \otimes |-\rangle$ & $\Psi_3=|-\rangle \otimes |+\rangle$  & $\Psi_4=|-\rangle \otimes |-\rangle$  \\
      \hline
      $(-1)^{N_x+N_y}$   & +1 & $-1$& +1&  $-1$\\
      $(-1)^{N_x+N_z}$   & +1 &  $-1$&  $-1$&  $+1$\\
      $(-1)^{N_y+N_z}$   & +1 & +1 &  $-1$& $-1$\\
       \hline
        \hline
    \end{tabular}
    \caption{The quantum number of parity on the four possible ground states}
    \label{tab:parity}
\end{table}
\end{center}
\end{widetext}

Now we show that under CBC, the ground state of the spin channels is non-degenerate. To preserve the $SO(3)$ symmetry, there are only two different CBCs: periodic(PBC) and anti-periodic(APBC), namely, $\psi_\alpha(x+L)=\psi_\alpha(x)$, and $\psi_\alpha(x+L)=-\psi_\alpha(x)$ for all $\alpha$.

 In terms of the bosonized fields, the parity operators can be written as 
\begin{eqnarray}
(-)^{N_x+N_y} =e^{i\sum_{\alpha=x,y} \phi_\alpha(L/2) -\phi_{\alpha}(-L/2)}  \\
(-)^{N_x+N_z} =e^{i\sum_{\alpha=x,z} \phi_\alpha(L/2) -\phi_{\alpha}(-L/2)} \\
(-)^{N_y+N_z} =e^{i\sum_{\alpha=y,z} \phi_\alpha(L/2) -\phi_{\alpha}(-L/2)}  
\end{eqnarray}
To have a uniform solution of $\Theta_1$ and $\Theta_2$(the  ground states are superpositions of uniform solutions), we must have
\begin{equation}
\theta_\alpha(x+L)=\theta_\alpha(x)+2n_\alpha \pi + \theta, \quad \alpha=x,y,z
\end{equation}
where $n_\alpha$ is integer and $\theta=0$ or $\pi$. Since $\Theta_1=\theta_x-\theta_y$, $\Theta_2=\theta_y-\theta_z$, they do not depend on the value of $\theta$. 

Consider the PBC case: $\psi_{R,L;\alpha}(x+L)=\psi_{R,L;\alpha}(x)$, and $\psi_{R,L;\alpha}(x)$ $\propto$ $ e^{-i(\pm \phi_\alpha -\theta_\alpha)}$, $\phi_\alpha$ must satisfy
\begin{eqnarray}
\phi_{\alpha}(x+L)&=&\phi_{\alpha}(x)+2n_{\alpha}\pi+\theta 
\end{eqnarray}
Note that $\theta$ can only take $0$ or $\pi$, otherwise, $\Psi_{R,L}$ can not satisfy the same boundary condition simultaneously. Now the aforementioned quantum number of the ground state should be $(-)^{N_x+N_y}$=1  $(-)^{N_x+N_z}$=1, and $(-)^{N_x+N_z}=1$, where we can see that only $|\Psi_1\rangle$ satisfies these quantum numbers. Therefore, if the system has the PBC, the ground state of spin channel is $|\Psi_1\rangle$=$(|0\rangle+|\pi\rangle) \otimes (|0\rangle+|\pi\rangle)$ and non-degenerate, which is also true for APBC. 
Therefore, the spin channel of the ground state is unique under CBCs.

\section{Order parameter}
\label{S2}
With three species of fermion, we can construct all the parameters using the Gell-mann matrice (in analog to the Pauli matrices for spin-1/2 fermions). The particle-hole channels are
\begin{align}
\Delta_{DW}^i=\Psi_R^\dagger \gamma^i \Psi_L
\end{align}
where $\tau^i$(i=1,2...8) are the eight Gell-mann matrices.  $\Psi_{R,L}=(\psi_{{R,L}x}, \psi_{{R,L}y}, \psi_{{R,L} z}^T$   where $\psi_{R,L\alpha}$ are the right and left moving fermion of the $\alpha$ species.     Similarly, the particle-particle channels are
\begin{align}
\Delta_{SC}^i=\Psi_R \gamma^i \Psi_L
\end{align}

Using the bosonization identity
$
\psi_{R\alpha}(x)=\frac{\eta_{\alpha}}{\sqrt{2\pi\alpha}} e^{ik_Fx} e^{-i(\phi_\alpha-\theta_\alpha)}$
 and $\psi_{L\alpha}(x)=\frac{\eta_{\alpha}}{\sqrt{2\pi\alpha}} e^{-ik_Fx} e^{-i(-\phi_\alpha-\theta_\alpha)}$, and choosing the eight Gell-mann matrices as
\begin{align}
&\gamma^1= \begin{pmatrix} 0& 1 & 0 \\ 1& 0 & 0 \\ 0&0&0 \end{pmatrix}, \gamma^2= \begin{pmatrix} 0& -i & 0 \\ i& 0 & 0 \\ 0&0&0 \end{pmatrix}, \, \gamma^3= \begin{pmatrix} 1& 0 & 0 \\ 0& -1 & 0 \\ 0&0&0 \end{pmatrix}, \nonumber \\
& \gamma^4= \begin{pmatrix} 0 & 0 & 1 \\ 0& 0 & 0 \\ 1&0&0 \end{pmatrix}, \gamma^5= \begin{pmatrix} 0&  0& -i \\ 0& 0 & 0 \\ i&0&0 \end{pmatrix}, \gamma^6= \begin{pmatrix} 0& 0 & 0 \\ 0& 0 & 1 \\ 0&1&0 \end{pmatrix}, \nonumber \\
& \gamma^7= \begin{pmatrix} 0&0  &0 \\ 0 & 0 & -i \\ 0&i&0 \end{pmatrix}, \gamma^8= \frac{1}{\sqrt{3}}\begin{pmatrix} 1& 0 & 0 \\ 0& 1 & 0 \\ 0&0&-2 \end{pmatrix}
\end{align}
 the order parameters become
 
 \begin{widetext}
\begin{align}
&\Delta_{DW}^1=\psi_{Rx}^\dagger \psi_{Ly}+\psi_{Ry}^\dagger \psi_{Lx}
=\frac{\eta_x\eta_y}{2\pi \alpha} e^{-i2k_Fx} 2i \text{sin}(\theta_y-\theta_x)e^{i(\phi_x+\phi_y)}\\
&\Delta_{SC}^1=\psi_{Rx} \psi_{Ly}+\psi_{Ry} \psi_{Lx} 
=\frac{\eta_x\eta_y}{2\pi \alpha}  2i \text{sin}(\phi_y-\phi_x)e^{i(\theta_x+\theta_y)}\\
&\Delta_{DW}^2=-i(\psi_{Rx}^\dagger(x) \psi_{Ly}-\psi_{Ry}^\dagger \psi_{Lx} )
=\frac{-2i\eta_x\eta_y}{2\pi \alpha} e^{-i2k_Fx}  \text{cos}(\theta_y-\theta_x)e^{i(\phi_x+\phi_y)}\\
&\Delta_{SC}^2=-i(\psi_{Rx} \psi_{Ly}-\psi_{Ry} \psi_{Lx}) 
=\frac{-2i\eta_x\eta_y}{2\pi \alpha}  \text{cos}(\phi_y-\phi_x)e^{i(\theta_x+\theta_y)}\\
&\Delta_{DW}^3=(\psi_{Rx}^\dagger \psi_{Lx}-\psi_{Ry}^\dagger \psi_{Ly} )
=\frac{1}{2\pi \alpha} e^{-i2k_Fx} [e^{i2\phi_x}-e^{i2\phi_y}]  \\
&\Delta_{SC}^3=\psi_{Rx} \psi_{Lx}-\psi_{Ry} \psi_{Ly} 
=\frac{1}{2\pi \alpha}  [e^{i2\theta_x}-e^{2i\theta_y}] \\
&\Delta_{DW}^4=\psi_{Rx}^\dagger \psi_{Lz}+\psi_{Rz}^\dagger \psi_{Lx}
=\frac{\eta_x\eta_z}{2\pi \alpha} e^{-i2k_Fx} 2i \text{sin}(\theta_z-\theta_x)e^{i(\phi_x+\phi_z)}\\
&\Delta_{SC}^4=\psi_{Rx} \psi_{Lz}+\psi_{Rz} \psi_{Lx} 
=\frac{\eta_x\eta_z}{2\pi \alpha}  2 \text{sin}(\phi_z-\phi_x)e^{i(\theta_x+\theta_z)}\\
&\Delta_{DW}^5=-i(\psi_{Rx}^\dagger \psi_{Lz}-\psi_{Rz}^\dagger \psi_{Lx} )
=\frac{-2i\eta_x\eta_z}{2\pi \alpha} e^{-i2k_Fx}  \text{cos}(\theta_z-\theta_x)e^{i(\phi_x+\phi_z)}\\
&\Delta_{SC}^5=-i(\psi_{Rx} \psi_{Lz}-\psi_{Rz} \psi_{Lx}) 
=\frac{-2i\eta_x\eta_z}{2\pi \alpha}  \text{cos}(\phi_z-\phi_x)e^{i(\theta_x+\theta_z)}\\
&\Delta_{DW}^6=\psi_{Ry}^\dagger \psi_{Lz}+\psi_{Rz}^\dagger \psi_{Ly}
=\frac{\eta_y\eta_z}{2\pi \alpha} e^{-i2k_Fx} 2i \text{sin}(\theta_z-\theta_y)e^{i(\phi_y+\phi_z)}\\
&\Delta_{SC}^6=\psi_{Ry} \psi_{Lz}+\psi_{Rz} \psi_{Ly} 
=\frac{\eta_y\eta_z}{2\pi \alpha}  2i \text{sin}(\phi_z-\phi_y)e^{i(\theta_y+\theta_z)}\\
&\Delta_{DW}^7=-i(\psi_{Ry}^\dagger \psi_{Lz}-\psi_{Rz}^\dagger\psi_{Ly} )
=\frac{-i\eta_y\eta_z}{2\pi \alpha} e^{-i2k_Fx} 2 \text{cos}(\theta_z-\theta_y)e^{i(\phi_y+\phi_z)}\\
&\Delta_{SC}^7=-i(\psi_{Ry} \psi_{Lz}-\psi_{Rz} \psi_{Ly}) 
=\frac{-i\eta_y\eta_z}{2\pi \alpha} 2 \text{sin}(\phi_z-\phi_y)e^{i(\theta_y+\theta_z)}\\
&\Delta_{DW}^8=\psi_{Rx}^\dagger(x) \psi_{Lx}(x)+\psi_{Ry}^\dagger \psi_{Ly} - 2\psi_{Rz}^\dagger \psi_{Lz}
=\frac{1}{2\pi \alpha} e^{-i2k_Fx} [e^{i2\phi_x}+e^{i2\phi_y}-2e^{i2\phi_z}]  \\
&\Delta_{SC}^8=\psi_{Rx} \psi_{Lx}+\psi_{Ry} \psi_{Ly}-2\psi_{Rz} \psi_{Lz}
=\frac{1}{2\pi \alpha}  [e^{i2\theta_x}+e^{2i\theta_y}-2e^{2i\theta_z}] \\
&\Delta_{DW}=\psi_{Rx}^\dagger \psi_{Lx}+\psi_{Ry}^\dagger \psi_{Ly} +\psi_{Rz}^\dagger \psi_{Lz}
=\frac{1}{2\pi \alpha} e^{-i2k_Fx} [e^{i2\phi_x}+e^{i2\phi_y}+e^{i2\phi_z}]  \\
&\Delta_{SC}=\psi_{Rx} \psi_{Lx}+\psi_{Ry} \psi_{Ly}+\psi_{Rz} \psi_{Lz} 
=\frac{1}{2\pi \alpha}  [e^{i2\theta_x}+e^{2i\theta_y}+e^{2i\theta_z}] 
\end{align}
\end{widetext}
where the last two quantities are related to the identity matrix which are both $SO(3)$ invariant.

Now we discuss the dominate correlation function for doped fermionic Haldane phase where $\phi_c$ is gapless while $\theta_{s_1}$ and $\theta_{s_2}$ are condensed (i.e $2\theta_x-2\theta_y$, $2\theta_y-2\theta_z$ are condensed, therefore $2\theta_y-2\theta_z$ is also condensed). Therefore $\phi_{s_1}, \phi_{s_2}$ is disordered and exponentially decaying, which indicates that $2\phi_x-2\phi_y$, $2\phi_y-2\phi_z$, $2\phi_y-2\phi_z$ are disordered and exponentially decaying.  Then $\Delta_{DW}^i$ and $\Delta_{SC}^i$($i\neq 3, 8$) are all exponentially decaying. Due to the $SO(3)$ symmetry, $\Delta_{DW}^{3,8}$ and $\Delta_{SC}^{3,8}$ are vanishing. Now only the $SO(3)$-singlet  superconductivity order parameter $\Delta_{SC}$ are power-law decaying: $\Delta_{SC}(r+r_0)\Delta_{SC}(r_0)\propto 1/r^{2/(3K_c)}$ .  

On the other hand, if the spin channels are gapped by condensing $2\phi_x-2\phi_y$ and $2\phi_y-2\phi_z$, 
then $\Delta_{SC}$ are exponentially decaying but $\Delta_{DW}$ are power-law: $\Delta_{DW}(r+r_0)\Delta_{DW}(r_0)\propto 1/r^{2K_c\over3}$. In addition to this $2k_F$ part contribution, there is also $zero$ momentum part, namely $\Delta_{DW}'=\sum_\alpha n_{\alpha, R}+n_{\alpha, L} \propto \partial \phi_c$, which decays faster than the $2k_F$ part since its power is $2K_c$. 

\section{Determination of the Phase Diagram}\label{S3}
\subsection{Central Charge and the Phase Boundary}
\begin{figure}
  \includegraphics[width=1\linewidth]{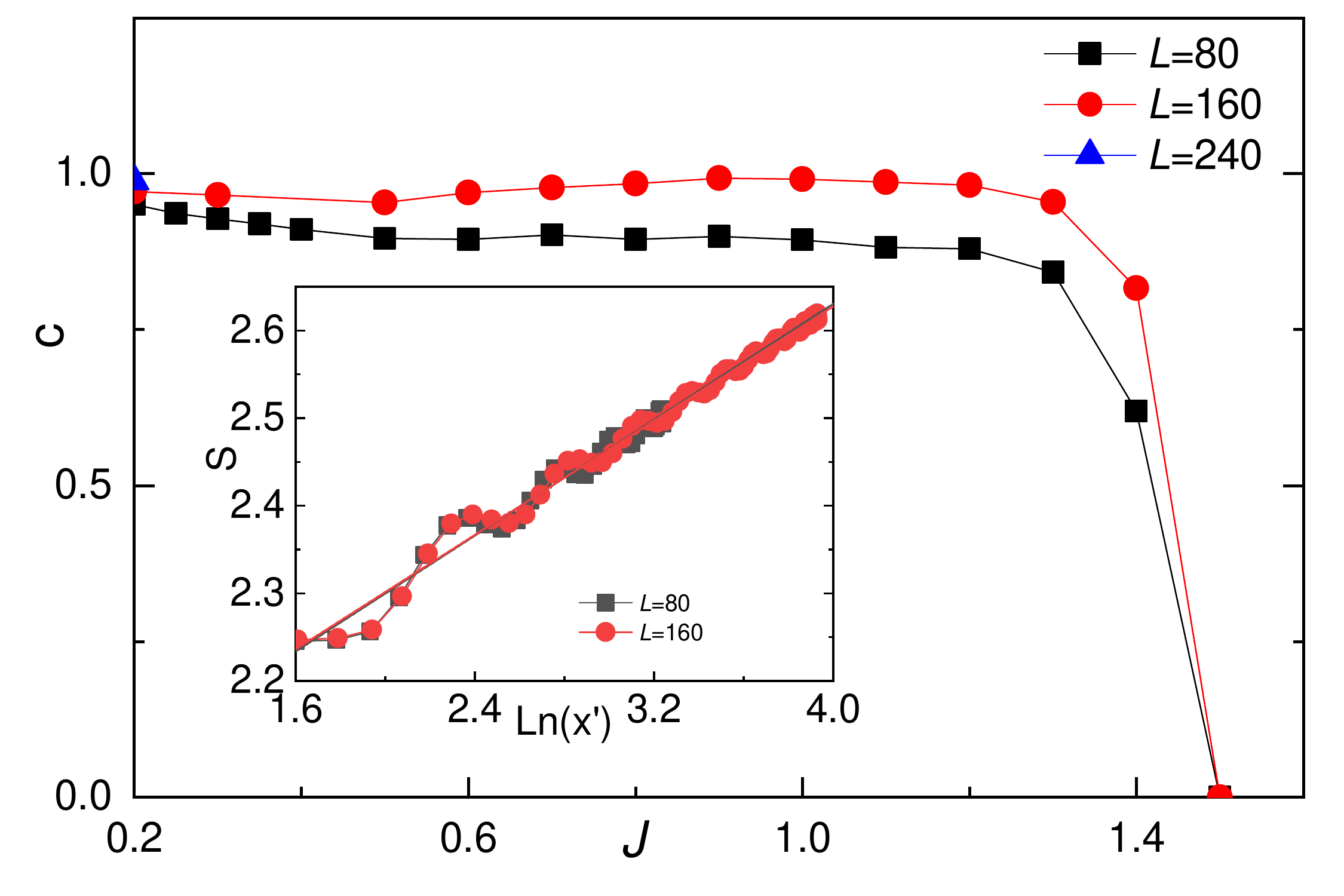}
  \caption{(Color online) The extracted central charge $c$ for the model in Eq.(2)(in the main text) at doping level $\delta=5\%$ and $U=4$ as a function of $J$. Inset: Von Neumann entanglement entropy $S$ at $U=4$ and $J=0.3$, where $x'=\frac{L}{\pi}\sin(\frac{x\pi}{L})$ and the linear line is the fit to $S=\frac{c}{6}\ln(x')+\tilde{c}$.}\label{Fig:EE_CC}
\end{figure}

Upon doping, if $J$ is not very large, the ground state of the system becomes superconducting with one gapless charge mode which is the $c=1$ Luttinger liquid. To confirm this, we calculate the von Neumann entanglement entropy. In the DMRG simulation, the central charge can be obtained by calculating the von Neumann entropy $S=-\rm Tr \rho ln \rho$, where $\rho$ is the reduced density matrix of a subsystem with length $x$. For critical system, it has been established\cite{Pasquale_2004} that $S(x)=\frac{c}{6}{\rm ln} (\frac{L}{\pi}\sin(\frac{x\pi}{L}))+\tilde{c}$ for open system, where $c$ is the central charge of the conformal field theory and $\tilde{c}$ denotes a model dependent constant. We use the central charge to determine the phase diagram(Fig.1 in the main text) in the main text. For typical parameters, our results, shown in Fig. \ref{Fig:EE_CC}, show that the central charge  is $c=1$ when $J<1.45$, consistent with one gapless charge mode while both spin modes are gapped(which only shift $\tilde c$ by some constant).

\subsection{Charge Density Profile and Phase Separation}

\begin{figure}[t]
    \centering
    \includegraphics[width=1\linewidth]{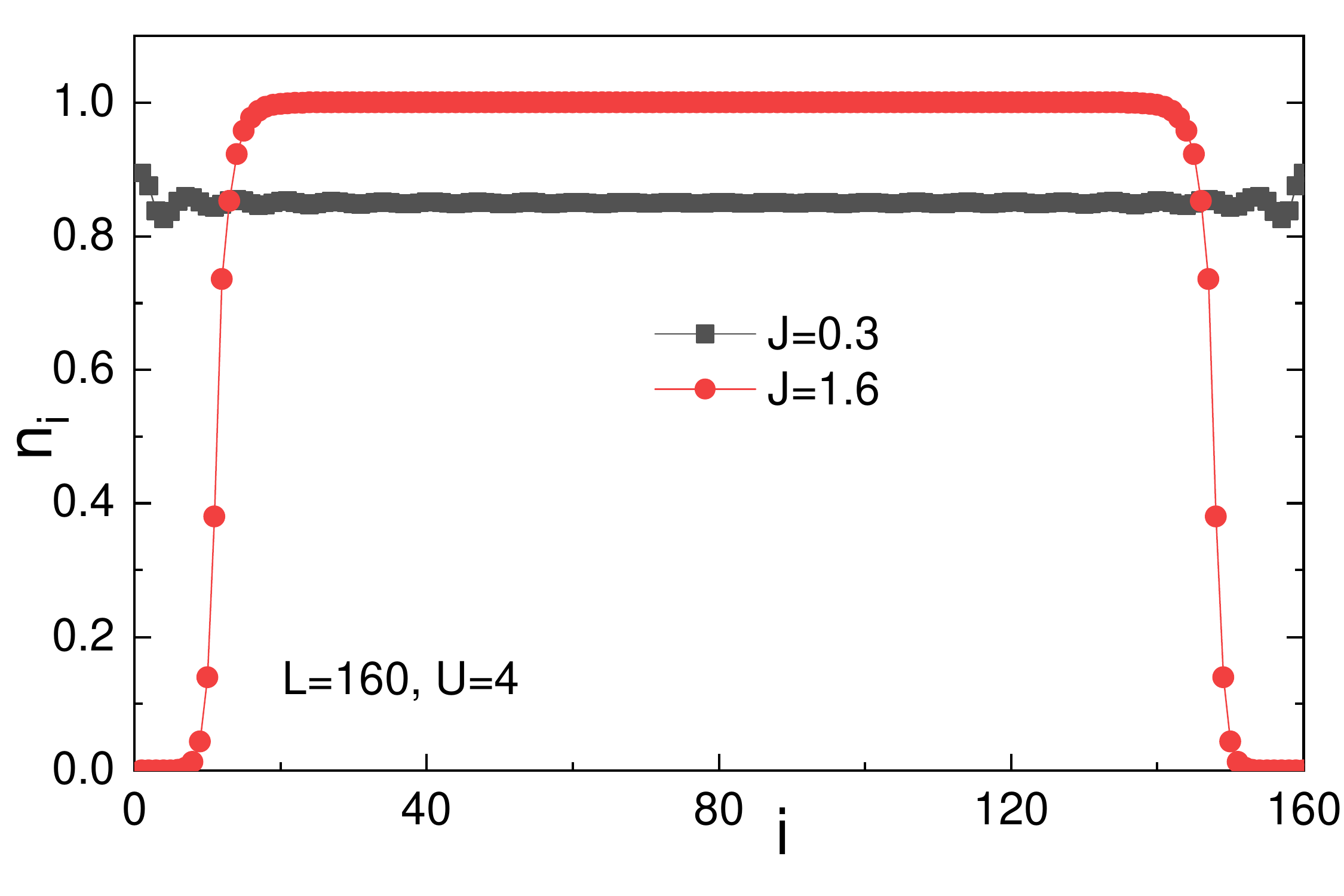}
    \caption{The density profile $n_i$ in the superconducting (black) and phase separation (red) regions. Here, $U=4$.}
    \label{density_profile}
\end{figure}

Here we will briefly discuss the phase separation (PS) region. As shown in Fig.\ref{density_profile} for $J/t=1.6$ in PS regime, the charge distribution is non homogeneous: while some parts of the system are full of particles, the rest parts of the system are empty without any particle, e.g., $n_i\sim 0$ close to the boundaries. This indicates that the doped holes are accumulated in the regions close to the boundaries, while the middle regime is at 1/3 filling which turns out to be in the fully gapped SPT phase.\cite{sq15} The PS phenomenon was also present in $t$-$J$ model(with projection)\cite{tj1/2, tjs}. For comparison, we also show the density profile in the SC regime in Fig.\ref{density_profile}, where a homogeneous charge distribution is seen despite the small charge oscillation caused by the  boundary effect.

Intuitively, this PS can be understood by comparing the energy gain from the hopping holes, and the energy cost by breaking the spin valence bonds. The energy gain from hopping holes is around $t$, while the energy cost of breaking the spin valence bonds is around $J$. To understand this, we turn to the undoped case where there is only one particle  per site if $U$ is large enough and hence  the  degree of freedom on every site is just spin-1. When this spin-1 is in the Haldane phase, the structure of this state can be viewed as the famous AKLT state\cite{AKLT} where the spin-1 are represented as two spin-1/2 that project on the spin-1 channel and two spin-1/2 from two nearest-neighbor sites form a spin singlet. The characteristic energy of such a spin singlet is around $J$. However, in the presence of holes, the on-site degree of freedom there is no longer the spin one which would cause the characteristic energy cost of $J$. Therefore, when $J$ is large enough, the energy cost of hopping hole overcomes the gain, which results in the phase separation. In other words, when the  interactions that drive the bSPT phase at certain filling are strong enough, they would repel the  charge fluctuation i.e. hole doping in it.  

\end{document}